\documentclass{llncs}

\usepackage{amsmath,amsfonts,amssymb}
\usepackage{ifthen}
\usepackage{pstricks, pst-plot, pst-node, pst-tree, pst-text}
\usepackage[dvips]{graphicx}
\usepackage{theorem}
\usepackage{enumerate}
\usepackage[T1]{fontenc}
\usepackage{times}
\usepackage{wrapfig}
\usepackage{url}

\newcommand{\acomplex}[1]{O(#1)}
\newcommand{\ecomplex}[1]{\Theta(#1)}
\newcommand{\lcomplex}[1]{\Omega(#1)}

\newcommand{\anode}{A}
\newcommand{\bnode}{B}
\newcommand{\nodeC}{C}
\newcommand{\clock}{t}
\newcommand{\localclock}[1]{\mathcal{C}_{#1}}

\newcommand{\clockl}{s_{\it{low}}}
\newcommand{\clockh}{s_{\it{high}}}
\newcommand{\messagedelay}{\delta}
\newcommand{\executiondelay}{\gamma}
\newcommand{\netconf}{C}

\newcommand{\activation}{\mathcal{A}_0}

\newcommand{\dead}[1]{d(#1)}

\newcommand{\netsize}{n}
\newcommand{\trnstrial}{k}

\newcommand{\wakecount}[1]{w(#1)}
\newcommand{\pre}[1]{\mathit{p}(#1)}
\newcommand{\suc}[1]{\mathit{s}(#1)}
\newcommand{\firstwake}{F}
\newcommand{\roundtrip}{R}
\newcommand{\anywake}{W}

\newcommand{\xpassive}{\mathsf{passive}}
\newcommand{\xactive}{\mathsf{active}}
\newcommand{\xleader}{\mathsf{leader}}
\newcommand{\xidle}{\mathsf{idle}}
\newcommand{\xmessage}[1]{\langle #1 \rangle}
\newcommand{\xhop}{\mathsf{hop}}

\newcommand{\prob}[1]{P(#1)}

\newcommand{\notact}{\alpha}
\newcommand{\notactv}{(1-\activation)}
\newcommand{\probrt}{\beta}

\newcommand{\pas}[1]{\rho(#1)}

\title{
  Asynchronous Bounded Expected Delay Networks
}

\author{
  Rena Bakhshi\inst{1}
\and
  J\"{o}rg Endrullis\inst{1}
\and
  Wan Fokkink\inst{1}
\and Jun Pang\inst{2}
}

\institute{
  Vrije Universiteit Amsterdam,
  Department of Computer Science,\\
  De Boelelaan 1081a,
  1081 HV Amsterdam,
  The Netherlands\\
  {\textbraceleft {\tt rbakhshi,joerg,wanf}\textbraceright{\tt@few.vu.nl}}
\and 
 Universit\'{e} du Luxembourg, Facult\'{e} des Sciences, de la Technologie et de la Communication\\
 6, rue Richard Coudenhove-Kalergi, L-1359 Luxembourg, Luxemburg\\
  {\tt jun.pang@uni.lu}
}
\begin{document}

\addtolength{\abovedisplayskip}{-1ex}
\addtolength{\belowdisplayskip}{-1ex}

\maketitle
\pagestyle{plain}

\begin{abstract}
We propose a probabilistic network
model, called asynchronous bounded expected delay (ABE), which
requires a known bound on the expected message delay. 
In ABE networks all asynchronous executions are possible,
but executions with extremely long delays are less probable.
Thus, the ABE model captures asynchrony that occurs in sensor networks and
ad-hoc networks.

At the example of an election algorithm, we show that the minimal assumptions 
of ABE networks are sufficient for the development of efficient algorithms.
For anonymous, unidirectional ABE rings of known size $\netsize$ we devise a probabilistic 
election algorithm having average message and time complexity $\acomplex{\netsize}$.
\end{abstract}

\section{Introduction}
\label{sec:introduction}
The two commonly used network models are synchronous and asynchronous.
In synchronous network all nodes
proceed simultaneously in global rounds.
While this model allows for efficient algorithms,
the assumptions are typically too strict for practical applications.
The (fully) asynchronous network model, on the other hand,
requires only that every message will eventually be delivered.
The assumptions of this model are generally too weak to study the time complexity of algorithms.

For the development of practically usable, efficient algorithms we need to find a golden mean 
between synchronous and asynchronous networks.
A possible approach is asynchronous bounded delay (ABD) networks~\cite{p144-Chou,Tel94}, 
where a fixed bound on the message delivery time is assumed.
Due to this assumption, ABD networks are generally closer to synchronous than to fully asynchronous networks.
The ABD model is a nice theoretical framework, but 
the assumption of a bounded message delay is often hard to satisfy in real-life networks.

We propose a probabilistic model, that we call {\it asynchronous bounded expected delay (ABE)} networks.
The ABE network model requires a known bound on the \emph{expected} message delay.
Thus, we strengthen the asynchronous network model with a minimal requirement
for analysing the time complexity of algorithms.

We elaborate on the advantage of ABE over the ABD network model.
A strong point in favour of the ABE network model is its probabilistic nature.
A probabilistic treatment of the message delay is for example
crucial to analyse protocols that employ a 
\emph{timeout mechanism} like in the TCP/IP protocol.
The assumption of the known bound on the expected message delay
allows for deriving a lower bound on the probability that
a message will arrive within a given time limit (before the timeout).
In contrast, it is impossible to evaluate algorithms
with timeout mechanism on the basis of the ABD network model.
The only assumption of the ABD network model is a fixed bound on the message delay.
If the timeout is greater or equal to this bound, then
all messages will arrive in time, rendering the timeout mechanism useless.
If the timeout is smaller than the bound on the message delay,
it is impossible to estimate how many messages will arrive before the timeout
(all messages could arrive after the timeout).

Moreover, messages sent via a physical channel may get lost or corrupted,
for example, due to material imperfections or signal inferences.
Since message transmission is unreliable, all we can settle for is a probability $p$ of successful transmission.
To ensure that a message arrives at its destination, it may need to be retransmitted (possibly multiple times) 
until the transmission has been successful. The number of necessary retransmissions for a message cannot be bounded:
with probability $(1-p)^\trnstrial$ a message requires more than $\trnstrial$ retransmissions,
and thus the message delay is unbounded.
While the message delay cannot be bounded,
from the probability $p$ we can derive the average number of needed retransmissions
and thereby the average message delay.
In particular, the average number of transmissions is 
$\trnstrial_{\mit{avg}} = \sum_{\trnstrial=0}^\infty (\trnstrial+1)\cdot (1-p)^\trnstrial\cdot p = \frac{1}{p}$.
Assuming that a successful transmission takes one time unit,
the average message delay is $\frac{1}{p}$ as well.
Assuming that we know the exact value of $p$
for physical channels may already be an unrealistic assumption.
However, frequently a lower bound $p_{\mit{low}} \le p$ on $p$ can be derived
from material properties in combination with the maximum strength of inference signals in a given environment.
Such a lower bound on $p$ is sufficient 
to derive an upper bound on the expected message delay,
and this suffices for ABE networks.

Although the assumptions of ABE networks are minimal,
it is possible to devise efficient algorithms.
We demonstrate this on an example of an election algorithm for
anonymous, unidirectional ABE rings
having (average) linear time as well as message complexity. 
So its efficiency is comparable to the most optimal election 
algorithms known for anonymous, synchronous rings~\cite{IR90}.

Election is the problem of determining a unique leader in a network,
in the sense that the leader (process or node) knows that it has been elected 
and the other processes know that they have not been elected. 
This is a fundamental problem in distributed computing and has many applications. 
For example, it is an important tool for breaking symmetry in a distributed system.
By choosing a process as the leader it is possible to execute centralised protocols 
in a decentralised environment.
Election can also be used to recover from token loss for token-based protocols,
by making the leader responsible for generating a new token when the current one is lost.
There exists a broad range of election algorithms; 
see e.g.\ the summary in~\cite{Tel94,Lyn96}.
These algorithms have different message complexity in the worst and/or average case.
Furthermore, they vary in communication mechanism ({\em asynchronous vs.\ synchronous}),
process names ({\em unique identities vs.\ anonymous}),
and network topology (e.g. {\em ring, tree, complete graph}).

Classical (deterministic) election algorithms are \cite{HS80,Fra82,Pet82,DKR82} for asynchronous 
rings with worst-case message complexity $O(n \log n)$,
and \cite{FredericksonL87} for synchronous unidirectional rings with worst-case 
message complexity $O(\netsize)$. Without additional assumptions, 
$\lcomplex{\netsize \log \netsize}$ is the lower bound on the average message 
complexity for asynchronous
rings~\cite{PKR84,Bod91}.

In an {\em anonymous network}, processes do not carry an identity.
As the number of processes in a network increases, 
it may become difficult to keep the identities of all processes distinct, 
or a network may accidentally assign the same identity to different processes.
In some situations, transmitting identities may be too expensive (e.g., FireWire bus, cf.~\cite{MS00}).
Since deterministic election is impossible in an anonymous network~\cite{Ang80}; 
randomisation is used to break the symmetry. 

If the network size is known, it is possible to construct a randomised  
election algorithm that terminates with probability one, e.g. \cite{IR81}. It exhibits infinite traces, 
but the probability that such an infinite trace is executed is zero. For unknown network size, 
the presence of an \emph{oracle} for leader detection is required, e.g.~\cite{FJ06}. 
In the absence of an oracle, there are several impossibility results for anonymous rings. 
No randomised algorithms can elect a leader in an anonymous ring if the ring size is known only within a factor of two \cite{Ang80}. 
Furthermore, algorithms for computing the ring size always have a positive probability of computing the wrong result \cite{Tel94}. 
Thus, there is no randomised algorithm that can elect a leader in an anonymous ring of unknown size.

We study the problem of election in anonymous, asynchronous rings.
For such rings, the best known election algorithms are \cite{IR81,FP06,HM98,BFPP08} with 
average message complexity $\ecomplex{\netsize\log \netsize}$.
Itai and Rodeh \cite{IR90} have proposed an algorithm for synchronous unidirectional anonymous rings; 
its average message complexity is $\acomplex{\netsize}$. The algorithm strongly depends on the notion 
of rounds, even though the idea of the activation parameter in this algorithm 
is similar to ours. The algorithm proceeds 
multiple elections, each of which consist of exactly $\netsize$ rounds. 
Both central ideas of the algorithm crucially depend on synchronous networks: the synchronous sending 
of the messages in the beginning of the election as well determining the number of active nodes by counting 
the messages passed through in $\netsize$ rounds. Thus, the similarities between our algorithm and that 
of Itai-Rodeh are limited to the activation parameter, which is a very natural choice in probabilistic 
systems. Hence, the Itai-Rodeh algorithm cannot be adapted for ABE networks; however, our 
algorithm works in synchronous networks.

\paragraph*{\bf Contribution and outline.} 
The ABD model assumes that there is a fixed bound on message delay. 
In Sec.~\ref{sec:abe} we propose a probabilistic network model, called
asynchronous bounded expected delay (ABE) networks,
that allows for an unbounded message delay,
and assumes a known bound on expected message delay.
This model is closer to the fully asynchronous network model than ABD networks.

For anonymous, unidirectional ABE rings of known ring size $\netsize$, we devise a probabilistic leader election 
algorithm with average message complexity $\acomplex{\netsize}$; see Sec.~\ref{sec:algorithm}. 
Previously, leader election algorithms with linear time and message complexity 
have only been known for network models with a strict bound on the message delay, i.e.,
synchronous networks and ABD networks.
In Sec.~\ref{sec:correctness} and~\ref{sec:complexity}, we prove the correctness, 
the average linear complexity of our algorithm, followed by an optimisation of the activation parameter.
The correctness proof and the complexity analysis are supported by an automated analysis with 
the probabilistic model checker PRISM in Sec.~\ref{sec:analysis}.

\section{Asynchronous Bounded Expected Delay Networks}
\label{sec:abe}
We introduce the model of ABE networks, 
which strengthens asynchronous networks with the 
assumption of a known bound on the expected message delay.
This strengthening allows one to analyse the (average) time complexity of algorithms.

At first glance it may appear superfluous
to consider a bound on the expected delay, instead of the expected delay itself.
We briefly motivate our choice.
First, when determining the expected delay for real-world networks, 
one needs to take into account
parameters such as material properties, environmental radiation, electromagnetic waves, etc.
Frequently, these values change over time, or cannot be calculated precisely.
Thus we have to cope with ranges for each of these parameters,
and consequently, the best we can deduce is an upper bound on the expected message delay.
Second, the links in a network are typically not homogeneous
and often have different expected delays.
Then the maximum of these delays can be chosen as an upper bound,
instead of having to deal with different delays for ever link.


\begin{definition}\normalfont
We call a network \emph{asynchronous bounded expected delay (ABE)} if the following holds:
\begin{enumerate}
  \item A bound $\messagedelay$ on the expected message delay (network latency) is known.
   \item Let $t$ be a real time. We assume that bounds $0 < \clockl \le \clockh$
   on the speed of the local clocks are known.
   That is, for every node $\anode$  the following holds for the local clock $\localclock{\anode}$: { }
   \(\clockl \cdot (\clock_2 - \clock_1) \le |\localclock{\anode}(\clock_2) - 
   \localclock{\anode}(\clock_1)| \le \clockh \cdot (\clock_2 - \clock_1) \;.\)
   \item A bound $\executiondelay$ on the expected time to process a local event is known.
\end{enumerate}
\end{definition}


In comparison with ABD networks, the ABE network model relieves the assumption
of a strict bound on the message delay.
The assumption is weakened to a bound on the expected message delay.
Thereby we obtain a probabilistic network model
which, as discussed above, covers a wide range of real-world networks
to which the ABD network model is not applicable.
For this reason, we advocate the model of ABE networks as a natural and useful extension of
the fully asynchronous network model.


\begin{example}
  The known upper bound on the expected message delay $\messagedelay$ allows for deriving
  a lower bound $p(t)$ on the probability that a message will be delivered 
  with a delay less or equal to $t$.
  From $\messagedelay \ge t \cdot (1-p(t))$
  it follows that $p(t) \ge 1- \messagedelay/t$
  for $t > \delta$.
  If $\messagedelay = 1$, 
  then $p(2) = 0.5$, $p(5) = 0.8$, etc.
  As a consequence we also obtain that long message delays are less probable (but possible).
\end{example}

To conclude this section, we discuss synchronisers for ABE networks.
A synchroniser is an algorithm to simulate a synchronous network on another network model.
A well-known impossibility result~\cite{4227} states that fully asynchronous networks cannot be synchronised 
with fewer than $\netsize$ messages per round (every node needs to send a message every round).
This of course destroys the message complexity when running synchronous algorithms in a fully 
asynchronous network.
The more efficient ABD synchroniser by Tel et al.~\cite{TKZ94} relies on knowledge of the bounded message delay.
As in fully asynchronous networks the message delay in ABE networks is unbounded
(although we assume a bound on the expected delay).
In a slogan: every execution of a fully asynchronous network
is also an execution of an ABE network. The difference is that huge message delays in ABE networks are less probable.
Hence, the impossibility result~\cite{4227} for fully asynchronous networks carries over
to ABE networks as follows:

\begin{corollary}
  ABE networks of size $\netsize$ cannot be synchronised with fewer than $\netsize$ messages per round.\qed
\end{corollary}

Hence, we cannot run synchronous algorithms in ABE networks
without losing the message complexity.
Although ABE networks are very close to fully asynchronous networks,
it turns out the model allows for the development of efficient algorithms.
We show this at the example of 
a surprisingly robust and efficient leader election algorithm with $\ecomplex{\netsize}$  average time and message complexity.

\section{Fast Leader Election with Bounded Expected Delay}
\label{sec:algorithm}

We present a leader election algorithm for anonymous, unidirectional ABE rings. 
The algorithm is parameterised by a \emph{base activation parameter} $\activation \in (0, 1)$. 
The order of messages is arbitrary between any pair of nodes. 
For simplicity, we assume that the expected time to process a local event is $0$, that is, $\executiondelay = 0$.
However, all results presented in this paper can be generalised straightforwardly for expected value $\executiondelay > 0$.

The algorithm presented below actually does not require continuous clocks.
It suffices that every node has a local timer \emph{ticking} once per (local) time unit.
Obviously such a discrete timer can be simulated using continuous clocks, 
thus, w.l.o.g.\ we assume that every node has a timer in the sequel.

During execution of the algorithm every node is in one of the following states: $\xidle$, $\xactive$, $\xpassive$ or $\xleader$; 
in the initial configuration all nodes are $\xidle$.
Moreover, every node $A$ stores a number $\dead{\anode}$, initially $1$.
The messages sent between the nodes are of the form $\xmessage{\xhop}$, 
where $\xhop \in \{1,\ldots,\netsize\}$ is the hop-counter of the message.
Every node $\anode$ executes the following algorithm:

\begin{itemize}
 \item \label{event:idlen} 
If $\anode$ is $\xidle$, then at every clock tick, with probability $1 - (1-\activation)^{\dead{\anode}}$, 
$\anode$ becomes $\xactive$, and in this case sends the message $\xmessage{1}$.
\item
If $\anode$ receives a message $\xmessage{\xhop}$, it sets $\dead{\anode} = \max(\dead{\anode},\xhop)$. 
In addition, depending on its current state, the following actions are taken:
\begin{enumerate}[(i)]
  \item \label{event:idle}
    If $\anode$ is $\xidle$, 
    then it becomes $\xpassive$ and sends the message $\xmessage{\dead{\anode}+1}$.
  \item \label{event:passive}
    If $\anode$ is $\xpassive$, 
    then it sends the message $\xmessage{\dead{\anode}+1}$.
  \item \label{event:active}
    If $\anode$ is $\xactive$, 
    then it becomes $\xleader$ if $\xhop = \netsize$, 
    and otherwise 
    it becomes $\xidle$, purging the message in both cases.
\end{enumerate}
\end{itemize}

In other words, messages travel along the ring and `knock out' all $\xidle$ nodes on their way.
That is, $\xidle$ and $\xpassive$ nodes \emph{forward} messages;
by forwarding a message, $\xidle$ nodes are turned $\xpassive$. 
If a message has knocked out an $\xidle$ node (at any point during its lifetime),
we refer to the message as \emph{knockout message}.
If a message hits an $\xactive$ node, then it is purged, and the active node becomes $\xidle$,
or is elected leader if $\xhop = \netsize$ (that is, if the node itself is originator of the message).
We say that a node has \emph{woken up} when it 
transits from the $\xidle$ to the $\xactive$ state.

The value $\dead{\anode}$ stores the highest received hop-count for every node.
It indicates that $\dead{\anode} - 1$ predecessors are $\xpassive$.
A higher value of $\dead{\anode}$ increases the probability that a node $\anode$ becomes $\xactive$.
By taking $1 - (1-\activation)^{\dead{\anode}}$ as wake-up probability for nodes $\anode$,
we achieve that the overall wake-up probability for all nodes stays constant over time.
This ensures that the algorithm has linear time and message complexity.
%
%

Note that we forward messages $\xmessage{\xhop}$ as $\xmessage{\dead{\anode}+1}$ instead of $\xmessage{\xhop+1}$.
This is used since the channels exhibit non-FIFO behaviour.
Consider the following scenario. A message $\xmessage{h}$ with high hop-count overtakes a message $\xmessage{\ell}$ with low hop-count,
and then $\xmessage{h}$ is purged by an $\xactive$ node $\anode$.
Then $\dead{\anode} = h$ and when $\xmessage{\ell}$ passes by $\anode$
its hop-count will be increased to $h+1$
(as if the overtaking would never have taken place).
Using $\xmessage{\xhop+1}$ instead of $\xmessage{\dead{\anode}+1}$,
there exist scenarios where all nodes are $\xpassive$ except for one $\xidle$ node $\bnode$,
and $\dead{\bnode} = 2$.
That is, $\dead{\bnode}$ is much lower than the actual number of $\xpassive$ predecessors of $\bnode$,
and as a consequence the overall wake-up probability would not stay constant over time.

We briefly elaborate on why the framework of ABE networks is essential for this leader election algorithm.
The bound on the expected delay is necessary for proving that the algorithm terminates with probability one,
and that the average time and message complexity are $\ecomplex{\netsize}$.
To the best of our knowledge, the algorithm is the first leader algorithm with the linear average time and 
message complexity in the settings of asynchronous anonymous rings without a fixed bound on the message delay.

\section{Correctness}
\label{sec:correctness}
Our leader election algorithm has terminated when all nodes are either $\xpassive$ or $\xleader$,
and no messages are in transit.
In this section we show that our leader election algorithm terminates with probability $1$, 
and upon termination always exactly one leader has been elected.
Our algorithm satisfies the following invariants:

\begin{lemma}\label{lem:dead}
  For every node $\anode$ at least $\dead{\anode}-1$ predecessors are $\xpassive$.
\end{lemma}
\begin{proof}
  Initially, the claim holds since $\dead{\anode} = 1$.
  Assume the claim would be wrong, then consider the first event invalidating the claim.
  By definition of the algorithm, $\dead{\anode}$ is the maximum hop-count
  of all messages received by $\anode$,
  and $\xpassive$ nodes stay $\xpassive$ forever.
  Therefore, we can restrict attention to the case that a node $\anode$ receives a message $\xmessage{x}$, 
  but fewer than $x-1$ predecessors of $\anode$ are $\xpassive$.
  The message $\xmessage{x}$ must have been sent by the predecessor $\bnode$ of $\anode$.
  The case of $\bnode$ being non-$\xpassive$ is trivial, since then it must have sent the message $\xmessage{1}$.
  If $\bnode$ is passive, then $x \le \dead{\bnode} + 1$, and 
  since the invariant holds for $\bnode$, $\dead{\bnode} - 1$ predecessors of $\bnode$ are $\xpassive$.
  Then $\dead{\bnode} = x - 1$ predecessors of $\anode$ are $\xpassive$.
  \qed
\end{proof}

\begin{lemma}\label{lem:oneleader}
  When a $\xleader$ node is elected, all other nodes are $\xpassive$.
\end{lemma}
\begin{proof}
 According to the algorithm, an $\xactive$ node $\anode$ is elected $\xleader$, when it receives the message $\xmessage{\netsize}$. Then $\dead{\anode} = \netsize$, so by Lemma~\ref{lem:dead}, all $\netsize-1$ other nodes are $\xpassive$.
  \qed
\end{proof}

\begin{lemma}\label{lem:message:active}
  There are always as many messages in the ring as $\xactive$ nodes.
\end{lemma}
\begin{proof} 
  Initially all nodes are $\xidle$ and the lemma holds.
  Let us consider all possible scenarios.
  If an $\xidle$ or $\xpassive$ node receives a message $\xmessage{\xhop}$,
  it will relay the message further. Thus, the number of messages and $\xactive$ nodes remains unchanged.
  If an $\xactive$ node receives a message, it changes its state to either $\xidle$ or $\xleader$. In 
  both cases, the message is purged. Thus, both messages and $\xactive$ nodes decrease by 1.
 If an $\xidle$ node becomes $\xactive$, it sends out a message. Thus, the number of messages 
 and the number of $\xactive$ nodes both increase by 1.%
 Finally, note that when an $\xactive$ node receives the message $\xmessage{\netsize}$, and becomes 
 $\xleader$, by Lemma~\ref{lem:oneleader}, all other nodes are passive, so that there are no other 
 messages in the ring.
  Hence, in all cases the invariant is preserved.
  \qed
\end{proof}
%
%
\begin{lemma}\label{lem:nonpassive}
  Always at least one node is not $\xpassive$.
\end{lemma}
\begin{proof}
  Only $\xidle$ nodes $\anode$ can become $\xpassive$, after receiving a message $\xmessage{\xhop}$.
  This message will be passed on as $\xmessage{\dead{\anode}+1}$. Hence, there is at least one message 
  travelling in the network, and, by Lemma~\ref{lem:message:active}, at least one $\xactive$ node in the 
  network.
  \qed
\end{proof}
Using these four invariants, we can show that our algorithm is correct.

\begin{theorem}\label{th:onelead}
Upon termination, exactly one leader has been elected.
\end{theorem}

\begin{proof}
Termination without elected leader is not possible, 
since by Lemma~\ref{lem:nonpassive}, there is always a non-$\xpassive$ node $\anode$
(if $\anode$ would be $\xactive$ there would be a message travelling by Lemma~\ref{lem:message:active}).
By Lemma~\ref{lem:oneleader}, if a leader has been elected, all other nodes are passive. 
Hence, upon termination there is a unique leader.
\qed
\end{proof}

\begin{theorem}\label{th:leader}
  The leader election algorithm terminates with probability one.
\end{theorem}
\begin{proof}
There exist only a finite number of network configurations $\netconf$.
For every non-terminated configuration $c\!\in\!\netconf$, there is a probability $\prob{c}\!>\! 0$ 
such that for every possible non-deterministic choice the probability
of the next scenario is at least $\prob{c}$:
\begin{itemize}
  \item 
  no $\xidle$ node becomes $\xactive$
  until all messages in the network are forwarded and received by $\xactive$ nodes
  (by Lemma~\ref{lem:message:active} there are as many messages as active nodes in the network);
  \item
  next, exactly one $\xidle$ node $\anode$ 
  (which exists by Lemma~\ref{lem:nonpassive}) becomes $\xactive$,
  and its message travels around the whole ring without any other node becoming $\xactive$.
  When $\anode$ receives its own message, it is elected leader
  and we have termination.
\end{itemize}
Taking $\zeta = \min \{\prob{c} \mid c \in \netconf \}$
we obtain that from every possible non-terminated configuration the probability of
termination is at least $\zeta > 0$.
Hence the algorithm terminates with probability one.
%
%
\qed
\end{proof}

\section{Complexity}
\label{sec:complexity}
In this section, we show that our algorithm has linear time and message complexity.
First, we give an intuition behind the linear complexity of the algorithm.
The crux is the choice of a suitable activation parameter $\activation(\netsize)$.
To achieve a linear complexity it suffices to take (the non-optimal)
$\activation(\netsize) = 1 - \sqrt[\netsize]{ (\netsize-1) / (\netsize+1) }$.
We briefly elaborate on this choice.
For simplicity we assume $\clockl = \clockh = 1$ and $\messagedelay = 1$ for this sketch.
If all nodes gamble once,
the probability that any of the nodes wakes up is
$1 - (1-\activation(\netsize))^n = 2/(n+1)$.
The bound $\messagedelay$ on the expected message delay
allows us to derive a bound $\roundtrip(n)$ on the 
expected time for a message to travel through the whole ring:
$\roundtrip(n) = \messagedelay \cdot \netsize = \netsize$.
Then the probability $\anywake(n)$ of any node waking up during $\roundtrip(n)$ time
is $1 - (1 - 2/(\netsize+1))^n$
which converges to:
 $\anywake(n) \to 1 - e^{-2}, 
 \text{ for }n \to \infty.$
This is the crucial observation yielding linear time and message complexity:
the time $\roundtrip(\netsize)$ for a round trip of a message is linear in $\netsize$,
and the probability $\anywake(n)$ of any node waking up during this time
is constant (i.e. asymptotically independent of $\netsize$).

Omitting the simplification $\clockl = \clockh = \messagedelay = 1$,
we have $\roundtrip(n) = \messagedelay \cdot \netsize = \netsize$, and
we employ the lower (upper) bound $\clockl$ ($\clockh$) on the clock speed
to derive a lower (upper) bound on the probability of any node waking 
up during the time.

For every node $\anode$ we define the \emph{activation count $\wakecount{\anode}$}
as the number of times this node has woken up. 
We now show the properties of the activation count.

\begin{lemma}\label{lem:between}
  Let $\anode$ and $\bnode$ be $\xactive$ or $\xidle$ nodes 
  such that the path from $\anode$ to $\bnode$ visits only passive nodes.
  Then the number of messages between $\anode$ and $\bnode$ is 
  \[\wakecount{\anode} - \wakecount{\bnode} + \xactive(\bnode)\]
  where $\xactive(\bnode)$ is $1$ if $\bnode$ is $\xactive$, and $0$ otherwise.
\end{lemma}
\begin{proof}
  Initially all nodes are idle and the lemma holds. That is, 
  $\xactive(\bnode) = 0$; moreover, $\wakecount{\anode} = \wakecount{\bnode} = 0$ 
  and there are no messages in the ring.
  For a node $\anode$ we denote  
  $\pre{\anode}$ and $\suc{\anode}$ 
  as the first $\xactive$ or $\xidle$ predecessor and successor of $\anode$, respectively.
  First, consider the case: $\anode = \bnode$.
  Then all nodes except for $\anode$ are $\xpassive$,
  and by Lemma~\ref{lem:message:active} there is a message in the ring iff $\anode$ is $\xactive$.
  For the remainder of the proof we assume $\anode \ne \bnode$, that is, $\anode \ne \suc{\anode}$
  and $\anode \ne \pre{\anode}$. Let us consider all possible events.

  If a $\xpassive$ node receives a message $\xmessage{\xhop}$, it relays the message further.
  Thus, the number of messages between $\xactive$ or $\xidle$ nodes remains unchanged.

  If an $\xidle$ node $\anode$ receives a message $\xmessage{\xhop}$,
  it becomes passive and relays the message further.
  Then, the number of messages between $\pre{\anode}$ and $\suc{\anode}$ is:
  \[(\wakecount{\pre{\anode}} - \wakecount{\anode} + 0) + (\wakecount{\anode} - \wakecount{\suc{\anode}} + \xactive(\suc{\anode}))
   = \wakecount{\pre{\anode}} - \wakecount{\suc{\anode}} + \xactive(\suc{\anode})\]

  If an $\xidle$ node $\anode$ becomes $\xactive$, it sends out a message. 
  The same holds for the number of messages between $\anode$ and $\suc{\anode}$.
  Then $\wakecount{\anode}$, and so 
  \(\wakecount{\anode} - \wakecount{\suc{\anode}} + \xactive(\suc{\anode})\),
  increases by $1$. 
  The number of messages between $\pre{\anode}$ and $\anode$ remains unchanged:
  \(\wakecount{\pre{\anode}} - \wakecount{\anode} + \xactive(\anode)\),
  both $\wakecount{\anode}$ and $\xactive(\anode)$ increase by $1$
  and thereby equal each other out.

  Finally, if an $\xactive$ node $\anode$ receives a message, it changes its state to either $\xidle$ or $\xleader$. 
  In both cases, the message is purged. 
  The number of messages between $\anode$ and $\suc{\anode}$ remains unchanged:
  \(\wakecount{\anode} - \wakecount{\suc{\anode}} + \xactive(\suc{\anode})\).
  The number of messages between $\pre{\anode}$ and $\anode$ decreases by $1$. The same holds for  
  \(\wakecount{\pre{\anode}} - \wakecount{\anode} + \xactive(\anode)\), 
  since $\xactive(\anode)$ decreases by $1$.
  Hence, in all cases the lemma holds.
  \qed
\end{proof}

\begin{lemma}\label{lem:da}
  Let $\anode$ and $\bnode$ be nodes
  such that $\anode$ is not $\xpassive$,
  the path from $\anode$ to $\bnode$ visits only passive nodes,
  and there are no knockout messages between $\anode$ and $\bnode$.
  Then the number of nodes between $\anode$ and $\bnode$
  (not counting $\bnode$) is $\dead{\bnode}-1$.
  (Note, if $\anode = \bnode$, we assume the path around the whole ring)
\end{lemma}

\begin{proof}
  Let $\pas{\nodeC}$ be the number of $\xpassive$ predecessors of $\nodeC$ (not counting $\nodeC$).
  We say that a node $\nodeC$ is \emph{informed} if $\dead{\nodeC}-1 = \pas{\nodeC}$.
  Likewise, a message $M$ with hop-count $h$ is called \emph{informed} 
  if exactly $h - 1$ nodes preceding $M$ are passive.

  We prove by induction over the number of events:
  for all nodes $\anode$ and $\bnode$ such that $\anode$ is not $\xpassive$,
  and the path from $\anode$ to $\bnode$ visits only passive nodes,
  either $\bnode$ is informed,
  or there is an informed knockout message between $\anode$ and $\bnode$.
  By Lemma~\ref{lem:dead} we have $\dead{\bnode}-1 \le \pas{\bnode}$.
  Initially all nodes are informed. We consider all possible events.
  If an informed (knockout) message is relayed by a $\xpassive$ node,
  the message stays informed and the node becomes informed.

  If a message knocks out an $\xidle$ node $\anode$, then either
  the node was informed and hence sends out an informed knockout message,
  or (by induction hypothesis) there must be an informed knockout message $M$ before $\anode$.
  In the latter case, if $M$ was the message received, then the node becomes informed.
  
  If an $\xactive$ node $\anode$ purges an informed knockout message, it becomes informed.
  
  Finally, if an $\xidle$ node $\anode$ becomes $\xactive$, it sends out a (non-knockout) message. 

  Hence, in all cases the lemma holds.
\end{proof}

As a direct consequence we obtain the following corollary stating
that the overall wakeup probability of all nodes in the ring stays basically constant.
\begin{corollary}\label{cor:allactive}
  Whenever there are no knockout messages travelling in the ring, 
{\abovedisplayskip1mm
\belowdisplayskip-1mm
  \begin{align*}
    1 - (1 - \activation)^{\netsize} = 1 - \prod_{\substack{ \anode\text{ is } \xidle \text{ or } \xleader}} (1 - \activation)^{\dead{\anode}}   
  \end{align*}}%
\end{corollary}
\begin{proof}
 There are two possible scenarios, where no knockout messages travel in the ring; 
 (i) there are only $\xidle$ and $\xpassive$ nodes, or (ii) node $\anode$ is 
 $\xleader$ and the other $\netsize -1 $ nodes are passive.
 Then we can divide the ring in chains of passive nodes followed by an $\xidle$ or $\xleader$ node $\anode$,
 and by Lemma~\ref{lem:da} we have $\dead{\anode}$ is the length of the chain plus $1$.
 As a consequence we obtain:
 $\sum_{\anode\text{ is } \xidle \text{ or } \xleader} \dead{\anode} = \netsize$. 
\end{proof}
In other words, the overall wakeup probability (for $\xidle$ and $\xactive$ nodes) may only decrease
as long as a knockout message travels through the ring.
This means that there is a message in the ring that has 
turned nodes from $\xidle$ to $\xpassive$
and did not yet encounter an active node (and hence did not complete a round trip).
As soon as this message is purged by an $\xactive$ node $\anode$,
the node updates its counter $\dead{\anode}$ representing the number of $\xpassive$ predecessors, 
thereby restoring the overall wakeup probability.

%

\begin{lemma}\label{lem:idle}
  If an $\xactive$ node $\anode$ with activation count $\wakecount{\anode}$
  receives a message while all other nodes have activation count smaller than 
  $\wakecount{\anode}$, then $\anode$ is elected as $\xleader$.
\end{lemma}
\begin{proof}
 Assume, towards a contradiction, that there are at least two $\xidle$ or 
 $\xpassive$ nodes in the ring. 
 By Lemma~\ref{lem:between} there are no messages between $\anode$ and its 
 first $\xactive$ or $\xidle$ predecessor, unequal to $\anode$. Hence, 
 $\anode$ can receive a message only if all other nodes are $\xpassive$.
 By Lemma~\ref{lem:da} it follows that after $\anode$ received this message,
 $\dead{\anode} - 1 = \netsize - 1$. Hence $\anode$ must have received the message $\xmessage{\netsize}$,
 and thus is elected leader.
\end{proof}
\begin{theorem}\label{thm:tlinear}
  The election algorithm has linear time and message complexity.
\end{theorem}
\begin{proof}
First, we prove the linear time complexity. 
Recall that the timer of every node ticks ones per unit of local time.
Thus, within $1 / \clockl$ global time the timer of every node ticks at least once.
By Lemma~\ref{cor:allactive}, $1 - \notactv^\netsize$ is a lower bound on the probability 
for at least one node waking up in $1 / \clockl$ time.
Hence, an upper bound $\firstwake$ on the expected time until the first node becomes $\xactive$ 
can be obtained as follows:
{\abovedisplayskip0in
\belowdisplayskip0in
\begin{align*}
\firstwake = \frac{1}{\clockl} \cdot \sum_{i=0}^{\infty} (i+1) \cdot \notactv^{\netsize i} \cdot (1 - \notactv^\netsize) 
= \frac{1}{\clockl} \cdot \frac{1}{1-\notactv^{\netsize}} 
\end{align*}}%
The expected time for a message to travel around the entire ring is $\roundtrip = \netsize \cdot \messagedelay$.
An upper bound $\anywake$ on the probability of any node
waking up during $\roundtrip$ time is $\anywake = 1 - \notactv^{\netsize \cdot \roundtrip \cdot \clockh}$.

The first node $\anode$ gets $\xactive$ after expected time $\firstwake$.
Then with probability $1 - \anywake$ no other nodes wakes
up while the message of $\anode$ travels around the ring, and in this case, 
after expected time $\roundtrip$ 
the node $\anode$ will be elected as $\xleader$.
With probability $\le \anywake$ another node wakes up while the message of $\anode$ makes a round trip.
Then after expected time $\roundtrip$
all non-$\xpassive$ nodes in the ring will have activation count at least one.
This can be seen as follows. The message of $\anode$ travels along the ring,
knocking out all nodes with activation count $< \wakecount{\anode}$,
until it is purged by an $\xactive$ node $\bnode$.
Then $\bnode$ must have become $\xactive$ after $\anode$,
and itself sent out a message. We continue with tracing this message of $\bnode$,
and successively apply the same reasoning
until we have finished one round trip, and are back to $\anode$.

In case another node has woken up while the message of $\anode$
was travelling around the ring, the described scenario repeats.
For this it is important to observe that by Lemma~\ref{cor:allactive}
the overall wakeup probability of all nodes stays constant.
As a consequence, after expected time $\firstwake$ a node $\bnode$ will
wake up, and get the highest activation count $\wakecount{\bnode}$.
By Lemma~\ref{lem:idle}, after expected time $\roundtrip$ 
either (i) $\bnode$ will be elected leader (probability $1-\anywake$),
or (ii) all non-$\xpassive$ nodes in the network have at least activation count 
$\wakecount{\bnode}$ (probability $\anywake$).
This leads to the upper bound on the expected time of termination 
\begin{align} \label{eq:exptime}
  \sum_{i=0}^\infty (i+1) \cdot (\firstwake + \roundtrip) \cdot \anywake^i \cdot (1-\anywake) 
   = \frac{1+\roundtrip \cdot \clockl - \notactv^{\netsize} \cdot \roundtrip \cdot \clockl}{(1-\notactv^{\netsize})\cdot (1-\anywake)\cdot \clockl}
\end{align}
For $\activation$ we choose the following expression depending on $\netsize$: 
\( \activation = 1 - \sqrt[\netsize]{ \frac{\netsize-1}{\netsize+1} }. \)
Using this activation parameter time complexity is linear with respect to the ring size $\netsize$. 
The derivation of this parameter will be explained in the next subsection.
Substituting the expression for $\activation$ into \eqref{eq:exptime}, we obtain:
\begin{align*}
\frac{\frac{1}{\clockl} +  (1 - \notactv^{\netsize}) \cdot  \netsize \cdot \messagedelay}{(1-\notactv^{\netsize}) \cdot \notactv^{\netsize^2 \cdot \messagedelay \cdot \clockh}} 
= \frac{\frac{1}{\clockl} + \frac{2}{\netsize+1} \cdot \netsize \cdot \messagedelay}{\frac{2}{\netsize+1} \cdot (\frac{\netsize-1}{\netsize+1})^{\netsize \cdot \messagedelay \cdot \clockh}} 
= \frac{ \frac{1}{\clockl} \cdot \frac{\netsize+1}{2} +  \netsize \cdot \messagedelay}{(\frac{\netsize-1}{\netsize+1})^{\netsize \cdot \messagedelay \cdot \clockh}}
\end{align*}
Noting that \( \lim_{n \rightarrow \infty} (\frac{\netsize-1}{\netsize+1})^{\netsize \cdot \messagedelay \cdot \clockh} = 
(\frac{1}{e^2})^{\messagedelay \cdot \clockh} \), we obtain that the time complexity 
(i.e. the expression above) is $\ecomplex{\netsize}$.

Now, we prove the linear message complexity. Since the time complexity is linear, 
here is $\xi > 0$ such that the average execution time of the algorithm $ \leq \xi \cdot \netsize$. 
A lower bound on the expected time until a node becomes $\xactive$ is:
$\firstwake_{\mit{low}} = \frac{1}{\clockh} \cdot \frac{1}{1-\notactv^{\netsize}}$.
Thus, the expected number of nodes ``waking up'' during the algorithm execution is 
$ \leq \frac{\xi \cdot \netsize}{\firstwake_{\mit{low}}}
= \xi \cdot \netsize \cdot \clockh \cdot (1-\notactv^{\netsize}) 
= \xi \cdot \clockh \cdot \frac{2 \cdot \netsize}{\netsize+1}.
$
Since every ``wake up'' gives rise to at most $\netsize$ messages (once around the ring) 
and by the expression above, the algorithm has linear message complexity.
\end{proof}

The concept of message complexity differs from bit complexity.
The message complexity refers to the expected number of messages until termination,
while \emph{bit complexity} is the total number of bits transmitted.
Our algorithms has $\ecomplex{\netsize}$ message complexity, 
and $\ecomplex{\netsize\cdot \log \netsize}$ bit complexity as the messages
are $\log \netsize$ in size.

\paragraph{Optimal Value for $\activation$}
The crux of the algorithm is the activation parameter $\activation$
influencing both the time and message complexity.
We optimise $\activation$ with respect to time complexity 
in dependence on the network size $\netsize$. 
For simplicity we assume $\messagedelay = 1$, $\clockl = \clockh = 1$.
We conclude this section with a discussion of the general case.

The value for $\activation$ (depending on $\netsize$) derived from the optimisation is applied in the proof of Theorem~\ref{thm:tlinear}.
This shows that the obtained activation parameter results
in linear time and message complexity for arbitrary values for $\messagedelay$, $\clockl$ and $\clockh$,
although the analysis does not take these values into account.

For the purpose of optimisation, 
we consider average-case scenarios instead of worst-case scenarios 
as used in the proof of linear time and message complexity (Theorem~\ref{thm:tlinear}).
Let $\notact = 1 - \activation$. 
Then, the average number of attempts before a first node becomes $\xactive$ is
\begin{eqnarray*}
(1&-&\notact^\netsize) \cdot 1 + (1 - \notact^\netsize) \cdot \notact^\netsize \cdot 2 + \ldots = (1 - \notact^\netsize) \cdot \sum_{i=0}^{\infty} \notact^{\netsize i} \cdot (i+1) 
= \frac{1}{1-\notact^{\netsize}}
\end{eqnarray*}
The probability $\probrt$ that a message of this first node completes its round-trip is
\(
\notact^{\netsize-1} \cdot \notact^{\netsize-2} \cdot \ldots \cdot \notact = \notact^{\frac{\netsize(\netsize-1)}{2}}
\).
since the expected time for a round-trip is $\netsize \cdot \messagedelay = \netsize$.
Note that in the proof of Theorem~\ref{thm:tlinear} we have used an upper bound on the worst case for $\probrt$,
namely $\notact^{n^2}$.

Thus, the average time required to elect a leader is
\begin{eqnarray*}
\probrt \cdot \frac{1}{1-\notact^\netsize} &+& (1 - \probrt) \cdot \probrt \cdot \frac{2}{1-\notact^\netsize} + (1-\probrt)^2 \cdot \probrt \cdot \frac{3}{1-\notact^\netsize} + \ldots 
= \frac{1}{\probrt \cdot (1-\notact^\netsize)}.
\end{eqnarray*}
\noindent We take $(1-\probrt)$ as the probability of fail trial for the message to make a round-trip.
%
We now derive an optimal value for $\activation$. 
Optimal here means that the average time to elect the leader is as low as
possible. 
That is, we minimise $\frac{1}{\probrt \cdot (1-\notact^\netsize)}$ by taking the derivative: 

 \[
(\netsize-1) \cdot \notact^{\frac{1}{2} \netsize (\netsize-1)} - (\netsize+1) \cdot \notact^{\frac{1}{2} \netsize (\netsize+1)} = 0
\quad\Longrightarrow\quad \activation = 1 - \sqrt[\netsize]{\frac{\netsize-1}{\netsize+1}}
\]
Note that for a large ring size $\netsize$, the optimal activation parameter $\activation$ converges to \( 1- \sqrt[\netsize^2]{\frac{1}{e^2}}. \)

\begin{figure}[!tb]
\begin{minipage}[t]{0.49\linewidth}
 \begin{center}
\scalebox{.5}{\includegraphics{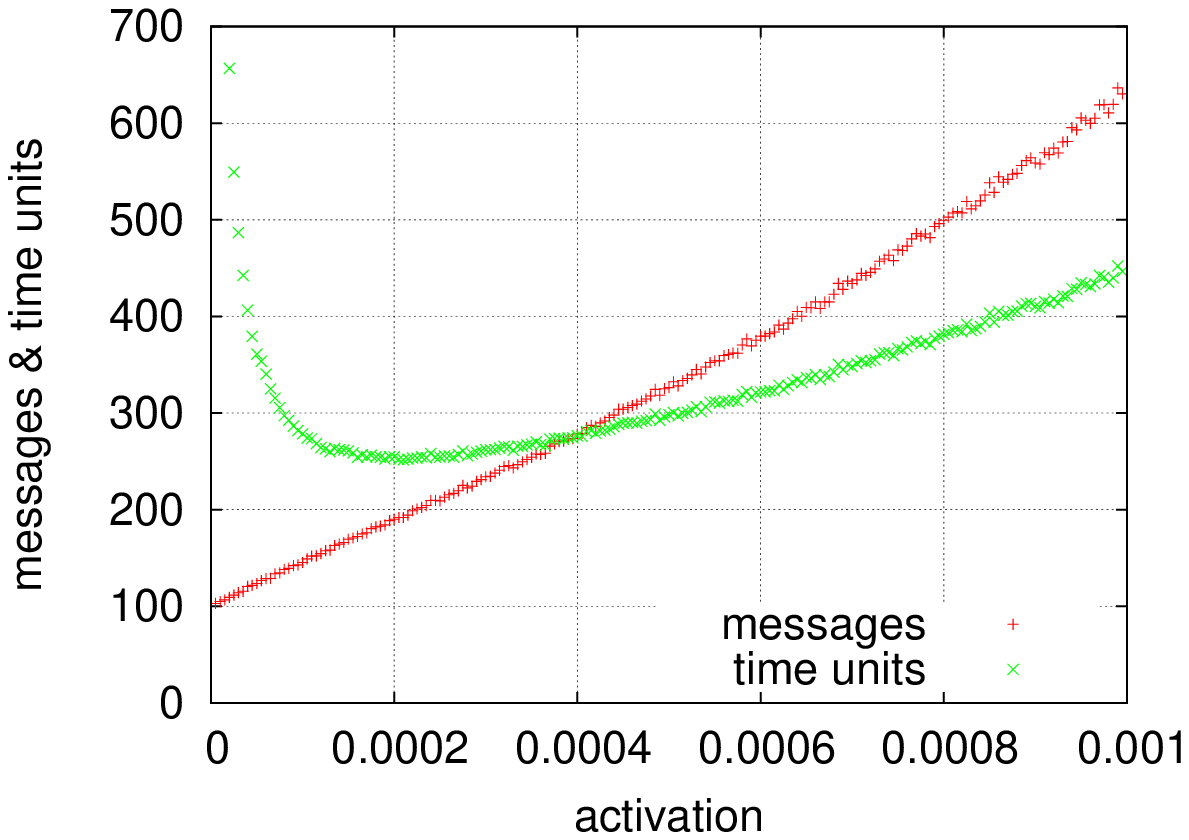}}
\end{center}
\end{minipage}
\begin{minipage}[t]{0.49\linewidth}
\begin{center}
\scalebox{.5}{\includegraphics{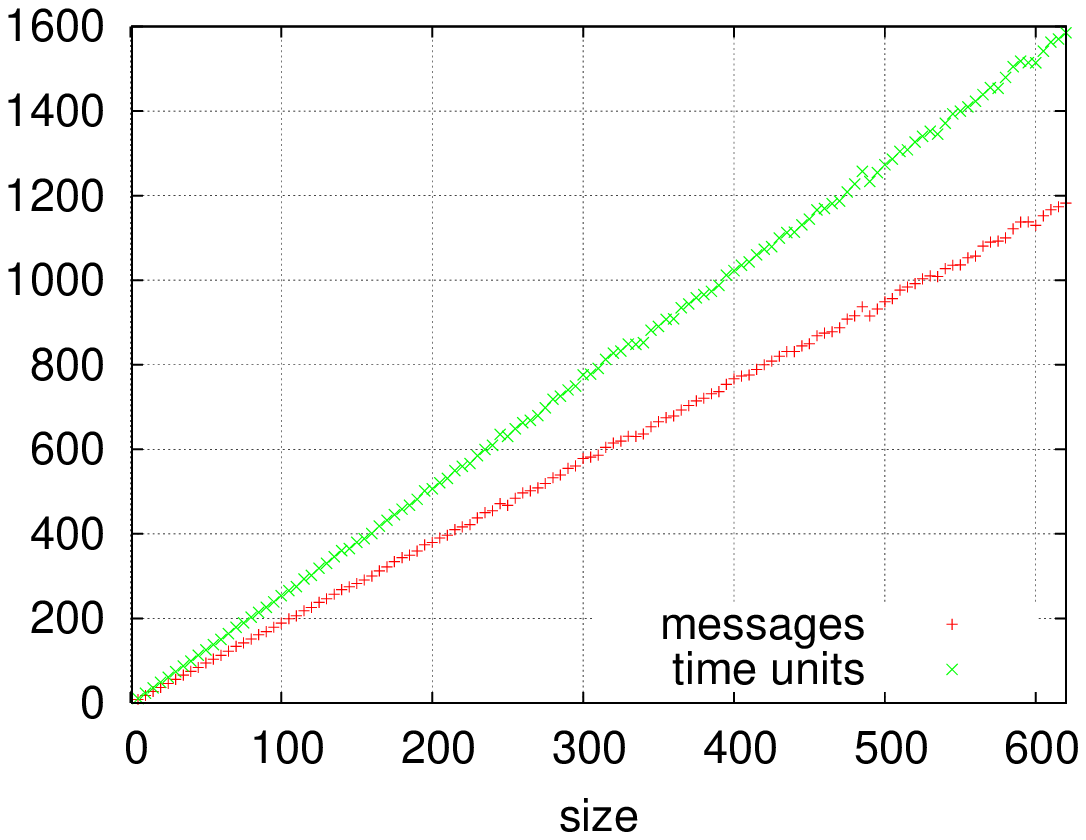}}
\end{center}
\end{minipage}
\caption{The different activation parameters $\activation$ for $100$ nodes (left). The optimal activation parameter for various ring size (right).}
\label{fig:tuning}
\end{figure}

We also simulated our algorithm in a round-based fashion, taking one global time unit as a measure of a round.
Fig.~\ref{fig:tuning} shows two set of experiments
(based on 5000 independent runs for each point of each curve).
The left graph of Fig.~\ref{fig:tuning} 
illustrates the impact of the activation parameter ($x$-axis) on the time and message complexity 
for a network with 100 nodes.
The experiment confirms the analytical derived optimal value for 
the activation parameter $\activation \approx 0.0002$ for $\netsize = 100$.

For the second set of experiments, we used the analytically derived optimal activation parameter. The right graph in Fig.~\ref{fig:tuning} shows the total number of messages and time units for the network size up to 620 nodes. We can see that our experimental results confirm the linear time and message complexity
for the optimal activation parameter $\activation$ depending on the network size $\netsize$.

We conclude this section with a brief discussion of the general case for
arbitrary $\messagedelay$, $\clockl$ and $\clockh$.
As already mentioned,
$\activation = 1 - \sqrt[\netsize]{(\netsize-1)/(\netsize+1)}$
results in linear time and message complexity for arbitrary values of these parameters.
However, the constant factor in $\ecomplex{n}$ can be high when optimising for the wrong parameters.
Note that, for the case $\clockl < \clockh$ we cannot speak about `the' optimal value for $\activation$
since we don't know the exact speed of every local clock.
However, we can optimise for the worst-case scenario,
then every real-world instance will only perform better.
That is,
to compute the average number of attempts before a first node becomes $\xactive$ we 
use $\clockl$ (slower ticking clocks imply lower wake-up probability),
and 
for probability of a message to complete its round-trip $\probrt$
we use $\clockh$ (faster ticking clocks imply worse probability) in combination with $\messagedelay$
(expected time for a round-trip is $n \cdot \messagedelay$).

\section{Automatic Analysis}
\label{sec:analysis}
To support the correctness proof and complexity analysis, we modelled our algorithm in the probabilistic model checker PRISM~\cite{HKNP06}. 
The properties like ``eventually exactly one leader is elected'' are expressed 
in the probabilistic temporal logic PCTL~\cite{HanJon94,BaiKwi98}. 
The algorithm is modelled in PRISM's state-based input language. 
Each process and each non-FIFO message channel is modelled as a module.
The message channels have the same size $\netsize$ as the ring network.
One global variable ${\cal A}$ is used for the activation probability.
The PRISM models and the properties in PCTL can be found at \\\hfill
\url{http://www.few.vu.nl/~rbakhshi/alg/onprism.tar.gz}.

For the model checking, we used PRISM (version 3.3). The option ``Use fairness" 
is turned on to restrict model checking only to fair paths.
In PRISM fairness is defined as follows: a path (in a computation tree) is \emph{fair}
if and only if for all states occurring infinitely often on the path,
each non-deterministic choice is taken infinitely often~\cite{RKGP04}.
It excludes infinite executions of the algorithm having probability $0.0$.
The other parameters in PRISM remain at default setting.

\begin{figure}[!t]
\begin{minipage}[c]{0.4\linewidth}
\begin{center}
\begin{tabular}{| r | r | r | r |}
\hline
\multicolumn{1}{|c|}{\hspace{3mm}$\netsize$} &
\multicolumn{1}{|c|}{States} &
\multicolumn{1}{|c|}{Transitions} &
\multicolumn{1}{|c|}{Min. Pr.}
\\
\hline
 3 & 397 & 921 & 1.0\\
\hline
 4 & 4,452 & 13,016 & 1.0\\
\hline
 5 & 50,659 & 180,070 & 1.0\\
\hline
 6 & 589,387 & 2,469,792  & 1.0\\
\hline
 7 & 6,980,446 & 33,683,860 & 1.0\\
\hline
\end{tabular}

\vspace*{3mm}
$\mathcal{A} \in \{ 0.1,0.2,\ldots,0.9 \}$ 
\end{center}
 \end{minipage}
 \begin{minipage}[c]{0.54\linewidth}
 \begin{center}
\scalebox{.57}{\includegraphics{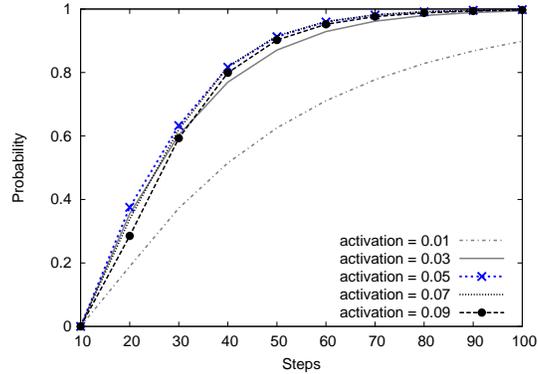}}
\end{center}
\end{minipage}
\caption{Model checking results in PRISM}
\label{fig:peran}
 \end{figure}

\paragraph{Correctness}
To verify the correctness, the algorithm is modelled as a 
Markov decision process, allowing both non-deterministic and probabilistic behaviour. 
In our experiments, we ask PRISM to compute the minimum probability that
eventually exactly one leader is elected. 
Fig.~\ref{fig:peran} (left) summarises our model checking results. 
The first columns in Fig.~\ref{fig:peran} (left) give the ring size $\netsize$ and
the number of states and transitions of the model. 
The second part shows the parameter ${\cal A}$
and the computed minimum probability for eventually electing a unique leader.
The minimum probabilities in all experiments are $1.0$,
which proves the correctness of the algorithm up to ring size seven.

\paragraph{Performance Analysis}
To compute the average-case complexity, the algorithm is modelled as a discrete-time Markov chain. 
For the first set of experiments, we used PRISM to compute the expected time to elect a leader for 
a fixed activation probability $\activation$ for $\netsize = 6$. 
That is, the probability that exactly one leader has been elected in the past
time units ${\it steps}$ (see Fig.\ref{fig:peran}, the right graph). 
The results correspond to our theoretical findings, namely, for $\netsize = 6$, the optimal value for the activation parameter is $\approx 0.0545$.



\bibliographystyle{splncs}
\bibliography{onb}

\begin{thebibliography}{10}

\bibitem{p144-Chou}
Chou, C.T., Cidon, I., Gopal, I.S., Zaks, S.:
\newblock Synchronizing asynchronous bounded delay networks.
\newblock IEEE Trans. on Communications \textbf{38}(2) (1990)  144--147

\bibitem{Tel94}
Tel, G.:
\newblock Introduction to Distributed Algorithms.
\newblock Cambridge University Press (2000) 2nd edition.

\bibitem{IR90}
Itai, A., Rodeh, M.:
\newblock Symmetry breaking in distributed networks.
\newblock Information and Computation \textbf{88}(1) (1990)  60--87

\bibitem{Lyn96}
Lynch, N.:
\newblock Distributed Algorithms.
\newblock Morgan Kaufmann Publishers (1996)

\bibitem{HS80}
Hirschberg, D., Sinclair, J.:
\newblock Decentralized extrema-finding in circular configurations of
  processors.
\newblock Comm. ACM \textbf{23}(11) (1980)  627--628

\bibitem{Fra82}
Franklin, R.:
\newblock On an improved algorithm for decentralized extrema finding in
  circular configurations of processors.
\newblock Commun. ACM \textbf{25}(5) (1982)  336--337

\bibitem{Pet82}
Peterson, G.:
\newblock An \({O}(n \log n)\) unidirectional algorithm for the circular
  extrema problem.
\newblock ACM Trans. Program. Lang. Syst. \textbf{4}(4) (1982)  758--762

\bibitem{DKR82}
Dolev, D., Klawe, M., Rodeh, M.:
\newblock An \({O}(n \log n)\) unidirectional algorithm for extrema finding in
  a circle.
\newblock J.\ of Algorithms \textbf{3}(3) (1982)  245--260

\bibitem{FredericksonL87}
Frederickson, G.N., Lynch, N.A.:
\newblock Electing a leader in a synchronous ring.
\newblock J. ACM \textbf{34}(1) (1987)  98--115

\bibitem{PKR84}
Pachl, J.K., Korach, E., Rotem, D.:
\newblock Lower bounds for distributed maximum-finding algorithms.
\newblock J. ACM \textbf{31}(4) (1984)  905--918

\bibitem{Bod91}
Bodlaender, H.:
\newblock New lower bound techniques for distributed leader finding and other
  problems on rings of processors.
\newblock Theor. Comput. Sci. \textbf{81} (1991)  237--256

\bibitem{MS00}
Maharaj, S., Shankland, C.:
\newblock A survey of formal methods applied to leader election in {IEEE} 1394.
\newblock J.\ of Universal Computer Science \textbf{6}(11) (2000)  1145--1163

\bibitem{Ang80}
Angluin, D.:
\newblock Local and global properties in networks of processors.
\newblock In: Proc.\ Symp.\ on Theory of Computing, ACM (1980)  82--93

\bibitem{IR81}
Itai, A., Rodeh, M.:
\newblock Symmetry breaking in distributive networks.
\newblock In: Proc.\ Symp.\ on Found. of Comput. Sci., IEEE (1981)  150--158

\bibitem{FJ06}
Fischer, M., Jiang, H.:
\newblock Self-stabilizing leader election in networks of finite-state
  anonymous agents.
\newblock In: Proc.\ Conf. on Principles of Distributed Systems. Volume 4305 of
  LNCS., Springer (2006)  395--409

\bibitem{FP06}
Fokkink, W., Pang, J.:
\newblock Variations on {I}tai-{R}odeh leader election for anonymous rings and
  their analysis in {PRISM}.
\newblock J.\ of Universal Computer Science \textbf{12}(8) (2006)  981--1006

\bibitem{HM98}
Higham, L., Myers, S.:
\newblock Self-stabilizing token circulation on anonymous message passing.
\newblock In: Proc.\ Conf.\ on Principles of Distributed Systems, Hermes (1998)
   115--128

\bibitem{BFPP08}
Bakhshi, R., Fokkink, W.J., Pang, J., {van de Pol}, J.C.:
\newblock Leader election in anonymous rings: Franklin goes probabilistic.
\newblock In: Proc.\ Conf.\ on Theoretical Computer Science. Volume 273 of
  IFIP., Springer (2008)  57--72

\bibitem{4227}
Awerbuch, B.:
\newblock Complexity of network synchronization.
\newblock J. ACM \textbf{32}(4) (1985)  804--823

\bibitem{TKZ94}
Tel, G., Korach, E., Zaks, S.:
\newblock {Synchronizing ABD networks}.
\newblock IEEE/ACM Trans. Netw. \textbf{2}(1) (1994)  66--69

\bibitem{HKNP06}
Hinton, A., Kwiatkowska, M., Norman, G., Parker, D.:
\newblock {PRISM}: A tool for automatic verification of probabilistic systems.
\newblock In: Proc.\ Conf.\ on Tools and Algorithms for the Construction and
  Analysis of Systems. Volume 3920 of LNCS., Springer (2006)  441--444

\bibitem{HanJon94}
Hansson, H., Jonsson, B.:
\newblock A logic for reasoning about time and reliability.
\newblock Formal Aspects of Computing \textbf{6}(5) (1994)  512--535

\bibitem{BaiKwi98}
Baier, C., Kwiatkowska, M.:
\newblock Model checking for a probabilistic branching time logic with
  fairness.
\newblock Distributed Computing \textbf{11}(3) (1998)  125--155

\bibitem{RKGP04}
Rutten, J., Kwiatkowska, M., Gethin, N., Parker, D.:
\newblock Mathematical Techniques for Analysing Concurrent and Probabilistic
  Systems.
\newblock American Mathematical Society (2004)

\end{thebibliography}

\end{document}